\documentclass[12pt]{article}
\usepackage{amsfonts}
\usepackage{amssymb}
\usepackage{amsbsy}
\usepackage{graphics}
\usepackage{color}
\usepackage{epsfig}

\input epsf.sty

\newlength{\TZ}
\newcommand{\BEQ}{\begin{equation}}     
\newcommand{\BEA}{\begin{eqnarray}}
\newcommand{\BD}{\begin{displaymath}}
\newcommand{\EEQ}{\end{equation}}       
\newcommand{\EEA}{\end{eqnarray}}
\newcommand{\ED}{\end{displaymath}}
\newcommand{\eps}{\varepsilon}          
\newcommand{\D}{{\rm d}}                
\newcommand{\II}{{\rm i}}               
\renewcommand{\Re}{{\rm Re\ }}          
\newcommand{\demi}{\frac{1}{2}}         
\newcommand{\wit}[1]{\widetilde{#1}}    
\newcommand{\wht}[1]{\widehat{#1}}      

\renewcommand{\vec}[1]{\boldsymbol{#1}} 

\newcommand{\R}{\mathbb{R}}
\newcommand{\C}{\mathbb{C}}

\catcode`\@=11
\def\numberbysection{\@addtoreset{equation}{section}
        \def\theequation{\thesection.\arabic{equation}}}
\numberbysection

\begin{document}

\begin{titlepage}

\vskip 1.5 cm
\begin{center}
{\LARGE \bf Non-local representations of the ageing algebra in higher dimensions}
\end{center}

\vskip 2.0 cm
\centerline{{\bf Stoimen Stoimenov}$^a$ and {\bf Malte Henkel}$^b$}
\vskip 0.5 cm
\centerline{$^a$ Institute of Nuclear Research and Nuclear Energy, Bulgarian Academy of Sciences,}
\centerline{72 Tsarigradsko chaussee, Blvd., BG -- 1784 Sofia, Bulgaria}
\vspace{0.5cm}
\centerline{$^b$ Groupe de Physique Statistique, D\'epartement de Physique de la Mati\`ere et des Mat\'eriaux,}
\centerline{Institut Jean Lamour (CNRS UMR 7198), Universit\'e de Lorraine Nancy,}
\centerline{B.P. 70239, F -- 54506 Vand{\oe}uvre l\`es Nancy Cedex, France}

\begin{abstract}
The ageing Lie algebra $\mathfrak{age}(d)$ and especially its local representations for
a dynamical exponent $z=2$
has played an important r\^ole in the description of systems undergoing simple ageing,
after a quench from a disordered
state to the low-temperature phase. Here, the construction of representations of
$\mathfrak{age}(d)$ for generic values
of $z$ is described for any space dimension $d>1$, generalising upon earlier results for
$d=1$. The mechanism for the
closure of the Lie algebra is explained. The Lie algebra generators contain higher-order
differential operators or the Riesz fractional derivative.
Co-variant two-time response functions are derived.
Some simple applications to exactly solvable models of
phase separation or interface growth with conserved dynamics are discussed.
\end{abstract}
\end{titlepage}

\setcounter{footnote}{0}
\section{ Introduction}

Understanding the cooperative properties of strongly interacting many-body problems continues to pose many challenges. Here, we
are interested in a particular class of non-equilibrium phenomena,
usually referred to as `{\em ageing}'. While first systematically
studied in structural glasses quenched from a molten state to below the
`glass-transition temperature' \cite{Stru78}, very similar
phenomena have been found in many other glassy and non-glassy systems far from equi\-li\-brium,
see e.g. \cite{Cugl02,Henkel10} for surveys
and further references.
Schematically, one may characterise ageing systems by (i) a slow relaxation dynamics,
(ii) absence of time-translation-invariance and (iii) dynamical scaling. Therefore,
if ageing is understood in this way, the study of the ageing of systems far from
equilibrium gives physically well-motivated setting
for the analysis of the consequences of dynamical scaling and the investigation
of possible generalisations and extensions.

In this work, we shall be interested in the dynamical symmetries of ageing systems.
We shall restrict throughout  to what is known as `{\em simple ageing}',
where the dynamics can described in terms
of a single time-dependent length $L(t)$ such that for large times
$L(t)\sim t^{1/z}$, which defines the {\em dynamical exponent} $z$. This growing length scale signals
a natural dynamical scaling in the long-time limit $t\to\infty$. It has been proposed that it might
be possible to enlarge this to a larger set of local scale transformation, to be referred to as
{\em `local scale-invariance'} ({\sc lsi}). The current state of {\sc lsi}-theory,
with its explicit predictions for two-time responses and
correlators, has been recently reviewed in detail in \cite{Henkel10}, where also the available numerical tests as well as the respective
exactly solved models are discussed.\footnote{One important consequence of {\sc lsi} is that for non-equilibrium scaling,
far from the stationary state, each scaling operator is characterised by {\em two} independent scaling exponents, labelled $x$ and $\xi$ in
eqs.~(\ref{eq:2}) and~(\ref{d1age}) below. While allowing for $\xi\ne 0$ is certainly required in certain exactly solved systems such as the
$1D$ Glauber-Ising quenched to temperature $T=0$ and certainly improves the fit with the numerical data,
it also appears that the agreement is not perfect. Very recently, it has been attempted to explain remaining
subtle differences in the scaling of the two-time responses $R(t,s)$
in several non-equilibrium models with $z<2$ terms of a `logarithmic' extension of {\sc lsi} \cite{Henkel12a,Henkel13a},
see \cite{Henkel13b} for a short review.} Here, we shall be concerned with certain algebraic techniques with the distant goal
to extend  known representations of {\sc lsi} algebras with dynamical exponents $z=2$ (or $z=1$) to more generic values.
As a first step in carrying out this programme, we shall study here certain non-local symmetries of some {\em linear} equations,
and shall leave the much more difficult case of genuinely non-linear equations for future work.

The analysis of the ageing of several simple magnetic systems,
without disorder nor frustrations, without any macroscopic conservation
law of the dynamics, and undergoing ageing when quenched
to a temperature $T<T_c$ below the critical temperature
$T_c>0$ is characterised by the dynamical exponent $z=2$ \cite{Bray94a}.
Furthermore, it has been suggested that the detailed scaling form of the
two-time correlators and responses can be obtained by an extension of
simple dynamical scaling with $z=2$ towards a larger Lie group
\cite{Henkel94,Henkel97,Henkel02}
of which the so-called `{\em ageing algebra}'
$\mathfrak{age}(d)= \left\langle X_{0,1}, Y^{(i)}_{\pm \demi},M_0, R_{ij}\right\rangle_{1\leq i<j\leq d}$
is the Lie algebra.
It can be defined in terms of its non-vanishing commutators  \cite{Henkel06a,Henkel06b}
\BEA
{} [X_n, Y^{(i)}_{m}]&=&\left( \frac{n}{2} -m\right) Y^{(i)}_{n+m},
\quad [X_n, X_{n'}]=(n-n')X_{n+n'}, \quad [Y^{(i)}_\demi, Y^{(j)}_{-\demi}]=\delta_{ij} M_0,
 \nonumber \\
{} [R_{ij}, R_{k\ell}] &=& \delta_{i\ell} R_{jk} + \delta_{jk}R_{i\ell} - \delta_{ik} R_{j\ell} - \delta_{j\ell}R_{ik},
\quad  [R_{ij}, Y^{(k)}_m]= \delta_{jk} Y^{(i)}_m- \delta_{ik} Y^{(j)}_m
\label{eq:agedcom}
\EEA
with $n,n'=0,1$, $m=\pm\demi$ and $1\leq i \leq  j\leq d$.
When acting on space-time coordinates $(t,\vec{r})$, a representation of
(\ref{eq:agedcom}) in terms of affine differential operators  is:
\BEA
X_0 &=& -t\partial_t - \demi (\vec{r}\cdot\partial_{\vec{r}}) - \frac{x}{2}
\hspace{3.15truecm} \mbox{\rm dilatation} \nonumber \\
X_1 &=& - t^2\partial_t - t (\vec{r}\cdot\partial_{\vec{r}}) - \frac{\cal M}{2} \vec{r}^2 - (x+\xi)t
\hspace{0.5truecm}\mbox{\rm special transformation}\nonumber \\
Y^{(i)}_{-\demi} &=& - \partial_{r_i}
\hspace{5.97truecm} \mbox{\rm space-translations} \label{eq:2} \\
Y^{(i)}_{\demi} &=& - t\partial_{r_i} - {\cal M}r_i
\hspace{4.52truecm} \mbox{\rm Galilei-transformation}\nonumber \\
M_0 &=& -{\cal M} \hspace{5.87truecm} \mbox{\rm phase shift} \nonumber\\
R_{ij} &=& r_i\partial_{r_j}-r_j\partial_{r_i}=-R_{ji}, \hspace{3.00truecm} \mbox{\rm rotations}.\nonumber
\EEA
together with the physical interpretation of these generators.
Clearly, the representation (\ref{eq:2}) has a dynamical exponent $z=2$
and acts {\em locally} on the space-time coordinates. Furthermore,
it generates a set of dynamical symmetries of the Schr\"odinger (or diffusion) equation:
\BEQ \label{eq:S}
{\cal S}\phi(t, \vec{r})=\left(2{\cal M} \partial_t + \frac{2{\cal M}}{t}(x+\xi-d/2)-\nabla^2_{\vec{r}}\right)\phi(t, \vec{r})=0,
\EEQ
in the sense that each of the generators of $\mathfrak{age}(d)$
maps a solution of (\ref{eq:S}) onto another solution. Herein,
$({\cal M}, x, \xi )$ is a triplet of constants which together characterise the solution
$\phi=\phi_{({\cal M},x,\xi)}$ of this
equation. Physically, $\cal M$
is interpreted as an inverse diffusion constant, whenever ${\cal M}\in\mathbb{R}$ or else as a
non-relativistic mass if ${\cal M}=\II m$ with $m\in\mathbb{R}$. Furthermore,
$x$ and $\xi$ are the two independent scaling dimensions
which characterise the field $\phi=\phi_{({\cal M},x,\xi)}$.\footnote{If one were to add the time-translations
$X_{-1}=-\partial_t$, one would obtain the embedding
$\mathfrak{age}(d)\subset \mathfrak{age}(d) \oplus \mathbb{C} X_{-1} \hookrightarrow \mathfrak{sch}(d)$
into the Schr\"odinger algebra $\mathfrak{sch}(d)$, with (\ref{eq:agedcom}) extended to $n=-1$.
In the representation (\ref{eq:2}) one then has $\xi=0$. Invariance of (\ref{eq:S}) implies further
that $x=d/2$, a fact essentially known to Jacobi and Lie in 1843 and 1881,
respectively and since then re-discovered many times.

Since time-translation-invariance is absent by definition in ageing systems
and hence a stationary state is never reached, the
new scaling dimension $\xi$ is a further universal characteristics of the relaxation process.
The transformation $\phi(t,\vec{r})=t^{\xi_{\phi}}\Phi(t,\vec{r})$ maps an
$\mathfrak{age}(d)$-quasi-primary scaling operator $\phi$ with the
scaling dimensions $(x_{\phi},\xi_{\phi})$ to the
$\mathfrak{sch}(d)$-quasi-primary scaling operator $\Phi$ with scaling dimensions
$(x_{\Phi}=x_{\phi}+2\xi_{\phi},\xi_{\Phi}=0)$ and confirmed for two-point functions \cite{Henkel06a}.
The relationship between three-point and higher correlators
is more involved \cite{Minic12}.}

For systems undergoing simple ageing with $z=2$, the extended local scale-invariance as described by the representation (\ref{eq:2}) of
$\mathfrak{age}(d)$ indeed seems to give an appropriate description, including several exactly solved examples where
$\xi\ne 0$ is required \cite{Henkel06a,Henkel10} -- the best-known example is the
$1D$ Glauber-Ising model quenched  to $T=0$.
The main prediction concerns the derivation of two-time (linear) responses
$R=R(t,s)=\left.\frac{\delta \langle \phi(t)\rangle}{\delta h(s)}\right|_{h=0}$
of the order parameter $\phi$ with respect to its conjugate magnetic field, from the assumption of co-variance
under the chosen representation of $\mathfrak{age}(d)$.
The derivation is based on the
Janssen-de Dominicis theory \cite{Janssen92} which permits to re-express responses
$R(t,s) = \left\langle \phi(t) \wit{\phi}(s)\right\rangle$
as a correlator of the order parameter $\phi=\phi_{({\cal M},x,\xi)}$ and the conjugate `response field'
$\wit{\phi}=\wit{\phi}_{(-{\cal M},\wit{x},\wit{\xi})}$ (see e.g. \cite{Janssen92,Tauber05,Tauber07}
for introductions and detailed references).
Co-variance under $\mathfrak{age}(d)$ then leads to a set of linear
partial differential equations for $R(t,s)$.\footnote{There is a systematic way to
`dualise' the parameter $\cal M$ which permits to further
extend $\mathfrak{age}(d)$-covariance to the covariance under a parabolic sub-algebra of the conformal algebra $\mathfrak{conf}(d+2)$ in
$d+2$ dimensions, which gives an algebraic way to derive the causality condition
$t>s$ necessary for the interpretation of the
co-variant two-point function as a physical response \cite{Henkel03a,Henkel12}.
In this dualised form, the non-relativistic AdS/CFT correspondence becomes evident, see e.g. \cite{Bala08,Minic08,Son08,Gray13}.}
In statistical physics, a common formulation uses a stochastic Langevin equation
\BEQ
\partial_t \phi(t,\vec{r}) = -D \frac{\delta {\cal H}[\phi]}{\delta \phi(t,\vec{r})} + \eta(t,\vec{r})
\EEQ
with a Ginzburg-Landau functional ${\cal H}$ and a centred gaussian noise
$\eta$ with a $\delta$-correlated second moment.
In the standard Janssen-de Dominicis formalism,
this is related to the equation of motion derived from the dynamic functional
\BEQ
{\cal J}[\wit{\phi},\phi] = \int \!\D t\D\vec{r}\: \left[ \wit{\phi}\left(2{\cal M}\partial_t - \Delta_{\vec{r}} \right)\phi
+ \wit{\phi}\, \frac{\delta{\cal V}[\phi]}{\delta \phi} - T {\wit{\phi}\,}^2 \right]
-\demi \int\!\D\vec{r}\D\vec{r}'\: \wit{\phi_0}(\vec{r}) a(\vec{r}-\vec{r}') \wit{\phi_0}(\vec{r}')
\EEQ
where $D^{-1}=2{\cal M}$, $\cal V$ is the `potential' part of the Ginzburg-Landau functional,
$a(\vec{r})$ the correlator in the initial state
$\wit{\phi_0}(\vec{r})=\wit{\phi}(0,\vec{r})$ and $T$
the temperature of the external heat bath. In the case of a quadratic potential
$\cal V$, this action produces two equations  of motion which both have a form similar to
(\ref{eq:2}), and with the above characterisations
of $\phi$ and $\wit{\phi}$ \cite{Picone04}. In this case, one has a decomposition
${\cal J}[\wit{\phi},\phi] = {\cal J}_0[\wit{\phi},\phi] + {\cal J}_b[\wit{\phi}]$ such that the `deterministic part'
${\cal J}_0$ is invariant under the action of the Galilei sub-algebra
$\mathfrak{gal}(d) = \left\langle Y^{(i)}_{\pm\demi}, M_0, R_{ij}\right\rangle_{1\leq i<j\leq d}$.
This implies the
Bargman super-selection rules \cite{Bargman54}. From this, it follows that \cite{Picone04,Henkel10}
\begin{enumerate}
\item the computation of all response functions can be exactly reduced to the computation of response in the `noiseless' theory
governed by the functional ${\cal J}_0$, whose averages are given by co-variance conditions under the dynamical symmetry algebra
\item all correlators can be exactly reduced to certain
integrals over higher response functions of the `noiseless' theory
\end{enumerate}
These calculations have been carried out for a long list of models undergoing simple ageing with $z=2$
\cite{Baumann05b,Durang09,Fuertes09,Henkel10}. In spite of many encouraging numerical results, there is not yet
a satisfactory analytical treatment available for a non-quadratic potential
${\cal V}$, where the main difficulty come from the
non-trivial co-factors of the scaling operators $\phi, \wit{\phi}$ under Galilei- and special transformations ($Y_{\demi}$, $X_1$).
A way around this problem might be to perform first a dualisation with respect to
$\cal M$, which converts the projective representation
(\ref{eq:2}) of the non-semi-simple Lie algebra $\mathfrak{age}(d)$
into a true representation of a sub-algebra of $\mathfrak{conf}(d+2)$
\cite{Stoimenov05,Stoimenov09}.

Can one extend this procedure, at least for linear stochastic Langevin
equations of motion, to arbitrary values of the dynamical exponent
$z$~? If we were to restrict to locally  realised algebras,
the recent classification of the non-relativistic limits of the
conformal algebra \cite{Duval09} would only admit the cases
(i) $z=1$: the conformal algebra $\mathfrak{conf}(d)$
or the conformal Galilean algebra $\mbox{\sc cga}(d)$ \cite{Havas78,Henkel97,Negro97},
eventually with the exotic central extension for $d=2$ \cite{Lukierski06}
(ii) $z=2$: the Schr\"odinger algebra and (iii) $z=\infty$; along with their sub-algebras.
Further examples can only be found when looking at non-local representation,
which means that we must go beyond the setting
for first-order linear (affine) differential operators in the generators and must look for different, non-local realisations of the
known abstract algebras. Some partial information is already available to serve as a guide:
\begin{enumerate}
\item the Galilei-invariance of the non-relativistic equation of motion
${\cal S}\phi=0$ should be kept, which means that algebraically
one should require
\BEQ
{} [Y_{\demi}^{(i)},Y_{-\demi}^{(j)}]=\delta_{ij} M_0,\quad [{\cal S},Y_{\pm\demi}^{(j)}]=\lambda_{\pm}^{(j)}{\cal S}\label{galalg}
\EEQ
where the scalars $\lambda_{\pm}^{(j)}$ are to be determined.
For example, in sub-representations where $M_0=0$, this can be extended to
a Schr\"odinger algebra representation for any value of $z$ \cite{Negro97},
which describes massless particles, but with a rather
trivial equation of motion, see below.
\item In the context of {\sc lsi},
different realisations of generalised symmetry algebras have been constructed by using
certain fractional derivatives \cite{Henkel02,Henkel07,Henkel10}.
The closure of these sets of generators can only be achieved by taking a
quotient with respect to a certain set of `physical' states.
Although this has been successfully applied to certain physical models \cite{Baumann07,Durang09}
the closing procedure is not completely determined and it is not clear how to obtain the group (finite) transformations.
\item There are many physically well-motivated systems,
such as the Kardar-Parisi-Zhang equation in  $1D$ with a dynamical exponent $z=3/2$
\cite{Kardar86}, diffusion-limited erosion \cite{Krug91} or else certain interacting particle-reaction models \cite{Spohn99,Popkov11}
with $z=1$ and a {\em non}-local equation of motion. One would like to
be able to discuss their dynamical symmetries in a systematic way, but these cases do not seem to be included in  the list of
known representations of known algebras.
\end{enumerate}

A distinct and potentially more promising method has been explored in \cite{Henkel11}. Therein, new non-local representations of
${\mathfrak{age}(1)}$ for an integer-valued dynamical exponent
$z=n\in\mathbb{N}$ were constructed. They read
\BEA
X_0 &=& -\frac{n}{2}t\partial_t - \demi r \partial_r - \frac{x}{2} \nonumber \\
X_1 &=&  \left(-\frac{n}{2}t^2\partial_t - t r \partial_r  - (x+\xi)t\right) \partial_r^{n-2} - \demi\mu r^2
\nonumber \\
Y_{-\demi} &=& - \partial_r \nonumber\\
Y_{\demi} &=& - t\partial_r^{n-1} - \mu r \nonumber \\
M_0 &=& -\mu. \label{d1age}
\EEA
The commutation relations (\ref{eq:agedcom}) are indeed satisfied, but with a notable exception, namely
\BEQ
\bigl[ X_1, Y_{\demi}\bigr] \:=\: \frac{n-2}{2} t^2 \partial_r^{n-3} {\cal S},
\EEQ
where the `Schr\"odinger equation' takes the form
\BEQ
\label{eq:schz1}
{\cal S}\psi(t,r) := \left( z \mu \partial_t - \partial_r^z +\frac{2\mu}{t}(x+\xi-\frac{z-1}{2})\right) \psi(t,r) = 0,
\EEQ
Therefore, the algebra is now closed only in the quotient space over solutions of eq.~(\ref{eq:schz1})
and in addition the representations (\ref{d1age}) act as dynamical symmetries of the same equation \cite{Henkel11}.\\

In this paper, we shall show how to generalise this construction (\ref{d1age})
to any spatial dimension $d\geq 1$ and to dynamical exponents $z\geq 2$; which will be important in order
to apply it to many physically relevant models. In contrast to the local case $z=2$,
the passage to $d>1$ is not trivial. We shall
describe this in section~2.
Co-variant two-point functions are computed from these non-local representations in section~3,
first for the case when $z$ is even (with special attention given to the case $z=4$)
and afterwards for generic values of $z$, when the Riesz fractional derivative will play
an important r\^ole. In section~4, we shall apply these results to some simple physical
models, namely the kinetic spherical model with a conserved order-parameter and quenched to $T=T_c$ and the Mullins-Herring (or Wolf-Villain)
equations of interface growth with mass conservation. The full space-time response are calculated from the non-local representations
of $\mathfrak{age}(d)$ and the results will be compared with the  known exact results \cite{Kissner92,Majumdar95,Sire04,Baumann07}. We
conclude in section~5.

\section{Non-local representations of $\mathfrak{age}(d)$}

As we shall now show, the non-local representations of $\mathfrak{age}(d)$ are quite different
for even and generic values of the dynamical exponent $z$. We shall discuss these cases separately.

\subsection{Representations of $\mathfrak{age}(d)$ with an even dynamical exponent}

Consider the following representations of
$\mathfrak{age}(d)$ for an even dynamical exponent $z=2n$:
\BEA
Y_{-1/2}^{(i)} &:=& - \partial_i\nonumber \\
Y_{+1/2}^{(i)} &:=& - t \partial_i \Delta^{n-1} - \mu r_i\nonumber\\
X_0 &:=& - n t\partial_t - \frac{1}{2} (\vec{r}\cdot\partial_{\vec{r}}) - \frac{x}{2}\nonumber\\
X_1 &:=& \left( - n t^2 \partial_t - t(\vec{r}\cdot\partial_{\vec{r}}) - (x+\xi)t \right) \Delta^{n-1} - \frac{\mu}{2} \vec{r}^2\nonumber\\
M_0 &:=& -\mu \label{eq:z2nalgebra}
\EEA
where  $\partial_i = \partial/\partial r_i$, $i=1,2,\ldots,d$ and
$\Delta=\Delta_{\vec{r}}=\partial_i\partial_i$ is the Laplacian. Upon setting $n=1$, one recovers (\ref{eq:2}).
The spatial rotations $R_{ij}$ are unchanged with respect to (\ref{eq:2}).
In order to check the commutation relations
(\ref{eq:agedcom}) we use the following relations:
\BEA
&& [\Delta^n, r_i]=2n\partial_i\Delta^{n-1}\nonumber\\
&& [\Delta^n, \vec{r}^2]=
4n\vec{r}\cdot\partial_{\vec{r}}\Delta^{n-1}+2n\left(d+2(n-1)\right)\Delta^{n-1}.
\label{eq:imprelations}
\EEA
Then the commutators (\ref{eq:agedcom}) are indeed satisfied, except the following:
\BEQ
[X_1, Y_{\demi}^{(i)}]= (n-1)t^2\partial_i\Delta^{n-2}{\cal S},
\label{eq:shellcond}
\EEQ
where
\BEQ
{\cal S}= 2n\mu\partial_t-\Delta^n+\frac{2\mu}{t}\left(x+\xi-\frac{d+2n-2}{2}\right),\nonumber
\EEQ
that is the generators (\ref{eq:z2nalgebra}) closed into the Lie algebra $\mathfrak{age}(d)$
only over functions from the solution space of the ``Schr\"odinger" equation
\BEQ
{\cal S}\Phi(t,\vec{r}) =
\left(2n\mu\partial_t-\Delta^n+\frac{2\mu}{t}\left(x+\xi-\frac{d+2n-2}{2}\right)\right)
\Phi(t,\vec{r})=0.\label{schrodinger}
\EEQ
The representation (\ref{eq:z2nalgebra}) acts as a dynamical symmetry of the same equation. Since
(\ref{schrodinger}) is linear, it is
enough to check the following commutators
\BEA && [{\cal S},Y^{(i)}_{-\demi}]=[{\cal S},Y^{(i)}_{\demi}]=[{\cal S},M_0]
=[{\cal S},R_{ij}]=0\nonumber\\
&& [{\cal S},X_{0}]=-n{\cal S}, \quad
[{\cal S},X_{1}]  = -2tn\Delta^{n-1}{\cal S}.\label{provesymmetry}
\EEA
Hence for any solution of ${\cal S}\Phi=0$, the infinitesimal change $\delta\Phi={\cal X}{\cal S}\Phi$
with a generator ${\cal X}$ chosen from (\ref{eq:z2nalgebra}) also solves the same equation.

\subsection{Representations of $\mathfrak{age}(d)$ with a generic dynamical exponent}

For generic values of the dynamical exponent $z$, an explicit representation of
$\mathfrak{age}(d)$ is more difficult to find
and turns out to be quite distinct from those found for an even-valued dynamical exponent. In particular,
when the dynamical exponent is an odd number, a generalisation of the
one-dimensional case in manner similar to $z$-even case
is possible only for a sub-algebra $\{X_0, X_1, Y_{\pm \demi}^i, M_0\}$
which does not include the rotations. In addition
this algebra is closed over solution of the system of $d$ equations,
which are not invariant under entire algebra. We shall
not develop this case here.

Instead, in order to treat the case with a generic value of $z$, recall first
the definition of the Riesz fractional derivative, see e.g.
\cite{Miller93,Samko93,Podlubny99,Mainardi07,Henkel10}
for further details.
This is a linear operator $\nabla^\alpha_{\vec{r}}$ acting as follows
\BEQ
\nabla^\alpha_{\vec{r}}f(\vec{r})=\II^\alpha\int_{\mathbb{R}^d}
\frac{\D \vec{k}}{(2\pi)^d}|\vec{k}|^{\alpha} \:
e^{\II \vec{r}\cdot\vec{k}} \,{\wht f}(k),\label{eq:derfract}
\EEQ
where the right-hand side as to be understood in a distribution sense and ${\wht f}(k)$
denotes the Fourier transform. Some elementary properties are, see e.g. \cite{Henkel10}
\BEA
&& \nabla^\alpha_{\vec{r}}\nabla^\beta_{\vec{r}}=\nabla^{\alpha+\beta}_{\vec{r}}\label{1prop}\\
&& \nabla^2_{\vec{r}}=\sum_{i=1}^d\partial_i^2=\Delta_{\vec{r}}\label{2prop}\\
&& [\nabla^\alpha_{\vec{r}}, r_i]=\alpha\partial_i\nabla^{\alpha-2}_{\vec{r}}\label{3prop}\\
&& [\nabla^\alpha_{\vec{r}}, \vec{r}^2]=2\alpha(\vec{r}\cdot
\partial_{\vec{r}})\nabla^{\alpha-2}_{\vec{r}}
+\alpha(d+\alpha-2)\nabla^{\alpha-2}_{\vec{r}}\label{4prop}\\
&&
\nabla^\alpha_{\mu\vec{r}}f(\mu\vec{r})=|\mu|^{-\alpha}\nabla^\alpha_{\vec{r}}f(\mu\vec{r})\label{5prop}
\EEA
We emphasise that the Riesz fractional derivative acts in many ways as a `square root' of the Laplacian, see eq.~(\ref{2prop}).

Now consider the generators
\newpage
\typeout{*** saut de page ***}
\BEA
Y_{-1/2}^{(i)} &:=& - \partial_i\nonumber \\
Y_{+1/2}^{(i)} &:=& - t \partial_i \nabla^{z-2}_{\vec{r}} - \mu r_i\nonumber\\
X_0 &:=& - \frac{z}{2} t\partial_t - \frac{1}{2} (\vec{r}\cdot\partial_{\vec{r}})
- \frac{x}{2}\nonumber\\
X_1 &:=& \left( -\frac{z}{2}t^2 \partial_t - t (\vec{r}\cdot\partial_{\vec{r}})
 - (x+\xi)t \right) \nabla^{z-2}_{\vec{r}} - \frac{\mu}{2} \vec{r}^2\nonumber\\
M_0 &:=& -\mu \nonumber \\
R_{ij} &:=& r_i \partial_j - r_j \partial_i = - R_{ji}\label{eq:z2mplus1algebra}
\EEA
Using the properties (\ref{3prop},\ref{4prop}) the commutators (\ref{eq:agedcom})
are seen to hold true, except for
\BEQ
[X_1, Y_{\demi}^{(i)}]=\frac{z-2}{2}t^2\partial_i\nabla^{z-4}_{\vec{r}}{\cal S},
\label{eq:shellcond2}
\EEQ
where
\BEQ
 {\cal S}= z\mu\partial_t-\nabla_{\vec{r}}^{z}+\frac{2\mu}{t}\left(x+\xi-\frac{d+z-2}{2}\right)\nonumber
\EEQ
Hence, the generators (\ref{eq:z2mplus1algebra}) close into a
Lie algebra on the quotient with respect to solutions of the equation
\BEQ
{\cal S}\Phi(t,\vec{r})
=\left(z\mu\partial_t-\nabla_{\vec{r}}^{z}+\frac{2\mu}{t}
\left(x+\xi-\frac{d+z-2}{2}\right)\right)\Phi(t,\vec{r})=0.
\label{fracsch}
\EEQ
As before, the representation (\ref{eq:z2mplus1algebra}) generates dynamical symmetries of the equation
(\ref{fracsch}) which can be seen from the commutators
\BEA && [{\cal S},Y^{(i)}_{-\demi}]=[{\cal S},Y^{(i)}_{\demi}]
=[{\cal S},M_0]=[{\cal S},R_{ij}]=0\nonumber\\
&& [{\cal S},X_{0}]=-\frac{z}{2}{\cal S}, \quad
[{\cal S},X_{1}]  = -zt\nabla_{\vec{r}}^{z-2}{\cal S}.\label{frprovesymmetry}
\EEA
Some comments are in order:
\begin{enumerate}
\item because of (\ref{2prop}),
the generators (\ref{eq:z2nalgebra}) for an even dynamical exponent $z=2n$ are included as well,
with the same invariant differential equation (\ref{fracsch}).
\item only if one adopts the correspondence $\partial_r \mapsto \nabla_{\vec{r}}$
between the partial derivative with respect to the $1D$ spatial coordinate $r$
and the Riesz fractional derivative $\nabla_{\vec{r}}$, one has a relationship between the
$1D$ representation (\ref{d1age}) and the representation (\ref{eq:z2mplus1algebra}) for $d>1$ dimensions.
\item the non-locality only enters into the
Galilei- and special transformations, generated by $Y^{(i)}_{\demi}$ and $X_1$.
\item certainly, the choice of the Riesz fractional derivative
is merely motivated by its convenient algebraic properties. It
is still unknown whether this is an appropriate choice  for the treatment of physical systems.
\end{enumerate}
Summarising, the representation of $\mathfrak{age}(d)$
proposed here explicitly uses generators acting non-locally
on space. In Fourier space, the form proposed here will lead to local,
but non-analytic generators.\footnote{Of course, in the limit $z\to 2$ one simply returns to the standard local
representation (\ref{eq:2}) of $\mathfrak{age}(d)$.} The special case of an even-valued dynamical
exponent appears to have rather special and possibly non-generic properties.

In addition, we have shown here that certain equations of motion can be said to posses the kind of non-local symmetry considered here,
at the price of somewhat weakening the meaning of the term
``dynamical symmetry'' with respect to how it had been used for the
cases where $z=1$ or $z=2$ admit generators acting locally.
It remains an open problem how to reformulate this kind of non-local
symmetry in terms of an invariance of a field-theoretic action.

\section{Co-variant two-point functions}

The two-point function
\BEQ
F(t_1, t_2, \vec{r}_1, \vec{r}_2)= \langle\Phi_1(t_1, \vec{r}_1)\Phi_2(t_2, \vec{r}_2)\rangle
\EEQ
can be derived from the covariance under representation (\ref{eq:z2mplus1algebra})
discussed in the previous section. Here
\BEQ
\vec{r}_1=(r_{11},...,r_{1i},...r_{1d}), \quad \vec{r}_2=(r_{21},...,r_{2i},...r_{2d}).
\EEQ
Later, we shall mainly concentrate on the case $z=4$,
relevant for the physical applications we have in mind (see section~4), but for the
moment $z$ will be left arbitrary. As usual in conformal or local scale-invariance,
one forms from (\ref{eq:z2mplus1algebra}) two-particle
operators and co-variance then gives the system ${\cal X}_i(2)F=0$.\\

\noindent {\bf 1.} The co-variance under space-translations requires:
  \BEQ
  Y^{(i)}_{-\demi}: (\partial_{r_{1i}}+\partial_{r_{2i}})F=(\partial_{r_i}-\partial_{r_i})F=0, \quad r_i=r_{1i}-r_{2i}.
  \EEQ
Letting $\vec{r}=(r_1,...,r_i,...r_d)$, it follows that $ F=F(t_1,t_2, \vec{r})$.\\

\noindent {\bf 2.} The covariance under a "mass"
transformations gives the Bargman super-selection rule $\mu_1+\mu_2=0$
and consequently only fields with opposite "masses" will give a
non-trivial result for the two-points function. It follows that
 \BEQ
 \Phi_1: (\mu; x_1, \xi_1), \quad\Phi_2: (-\mu; x_2, \xi_2),\nonumber
 \EEQ
where $x_1, x_2$ and $\xi_1, \xi_2$
are the scaling dimensions which characterise the corresponding non-stationary
quasi-primary scaling operators. Consequently, the scaling operator
$\Phi_1$ with a positive mass $\mu>0$ will be associated with
the order parameter $\phi$ whereas the scaling operator $\Phi_2$ with a negative mass
$-\mu<0$ is interpreted as the corresponding
response operator $\wit{\phi}$ in the context of Janssen-de Dominic theory, see section~1. \\

\noindent {\bf 3.}  Next we rewrite the two-point function as
\BEQ
F = F(\tau,v,\vec{r}) \;\; , \;\; \tau := t_1 - t_2 \;\; , \;\; v := t_1 / t_2,
\EEQ
and obtain from the three co-variance conditions
$Y^{(i)}_{\demi}F=0$ for $i=1,\ldots,d$, $X_0F=0$ and $X_1 F=0$ the equations
(with an implicit sum over repeated indices $j=1,\ldots,d$)
\BEA
\hspace{-1.0truecm}\left[ -  \tau\partial_{r_i}\nabla^{z-2}_{\vec{r}} - \mu_1 r_i \right] F &=& 0
\label{4.4} \\
\hspace{-1.0truecm}\left[ - \frac{z}{2}\tau\partial_\tau - \frac{1}{2} r_j \partial_{r_j} - \frac{x_1 + x_2}{2} \right] F &=& 0
\label{4.5} \\
\hspace{-1.0truecm}\left[\left(- \frac{z}{2}\tau^2\frac{v+1}{v-1}
\partial_\tau - \frac{z}{2}\tau v \partial_v- \frac{\tau v}{v-1} r_j \partial_{r_j}
- (x_1+\xi_1)\frac{\tau v}{v-1} - (x_2+\xi_2)\frac{\tau}{v-1}\right)\nabla^{z-2}_{\vec{r}}
\right. & & \nonumber \\
 \left. -\frac{\mu_1}{2} r_i^2 \right] F &=& 0 \label{4.6}
\EEA
Acting with $\nabla^{z-2}_{\vec{r}}$ on (\ref{4.5}), the difficult eq.~(\ref{4.6}) can be simplified to
\BEQ
\nabla^{z-2}_{\vec{r}} \left( \frac{z}{2} v \partial_v + \frac{v}{v-1}\left( x_1 - x_2 +2\xi_1 -2n+2\right) +
\frac{1}{v-1}\left( x_2 - x_1 +2\xi_2 -z+2\right) \right) F = 0 \label{4.7}
\EEQ
In analogy with the one-dimensional case \cite{Henkel11},
it is clear that each of the equations (\ref{4.4},\ref{4.5},\ref{4.7}) will fix the dependence of $F=F(\tau,v,\vec{r})$ on one of
its variables. In fact, since the variable $v$ does not enter explicitly into eqs.~(\ref{4.4},\ref{4.5}),
these two alone will completely fix
the dependence of $F(\tau,v,\vec{r})$ on $\tau$ and $\vec{r}$ such that the dependence on $v$ factorises.
Then the $(z-2)$-fold (fractional) derivatives in
(\ref{4.7}) can be dropped. We also checked that the closure condition of our representation is
automatically satisfied, as it should be,
and as seen for $d=1$ before \cite{Henkel11}. Isolating the usual power-law pre-factor
which only depends on the single time $t_2$, we find the following reduced form
\BEQ \label{4.8}
F  = t_2^{-(x_1+x_2)/z}\: (v-1)^{-\frac{2}{z}[ (x_1+x_2)/2+\xi_1+\xi_2-z+2]} \:
v^{-\frac{1}{z}[ x_2-x_1+2\xi_2-z+2]}\: f\left( \tau, \vec{r}\right).
\EEQ
where the last function $f(\tau,\vec{r})$ has to be determined from eqs.~(\ref{4.4},\ref{4.5}).

\noindent {\bf 4.} The co-variance condition under each rotation $R_{ij}$ gives,
for {\em scalar} quasi-primary operators $\Phi_{1,2}$
\BEQ
\left(r_i\partial_{r_j}-r_j\partial_{r_i}\right)F=0\label{eq:rotation}
\EEQ
and leads to $f=f(\tau,u)$, with $u=\vec{r}^2$ as expected.

Combining this with dynamical scaling eq.~(\ref{4.5}), one has simply
\BEQ
f = f\left( u^{z/2} \tau^{-1} \right)
\EEQ
and the form of the last remaining function $f$
has to be found from Galilei-covariance, as given by (\ref{4.5}).

This is the technically most demanding part of the calculation.
For the sake of computational simplicity,
we shall distinguish the cases of an even dynamical exponent $z=2n$ (in particular,
we shall describe below an explicit application of the case $z=4$)
before we consider the case of generic values of $z$.

\subsection{Even dynamical exponent $z=2n$}

Using the notation $p := u^{z/2} \tau^{-1}$, Galilei-covariance gives, for each $i=1,\ldots,d$
\BEQ
(\tau\partial_{r_i}\Delta^{n-1}_{\vec{r}}+\mu r_i)f(\tau, \vec{r})
= r_i\left(2\tau\partial_u\Delta^{n-1}_u+\mu\right)f(\tau,u)
= r_i\left( (2n)^np^\frac{n-1}{n}\partial_p\Delta^{n-1}_p +\mu\right)f(p)=0.\label{eq:galilei}
\EEQ
Here we use the following notations for the various forms of the Laplacian
\BEA
\Delta_{\vec{r}} & = & \sum_{i=1}^d\frac{\partial^2}{\partial r_i^2} =: \Delta_u = 2d\partial_u+4u\partial^2_u\nonumber\\
& = & \frac{2n}{\tau^{1/n}}\left((d+2(n-1))p^\frac{n-1}{n}\partial_p+2n p^\frac{2n-1}{n}\partial_p^2\right)=:\frac{2n}{\tau^{1/n}}
\Delta_p.\label{laplas}
\EEA
In contrast to $\Delta_{\vec{r}}^{n-1}$ which is a formal power series of $\Delta_{\vec{r}}$, the explicit
forms of $\Delta_u^{n-1}$ and $\Delta_p^{n-1}$ must be calculated for any specific value of $n$. For example
\BEQ
 \Delta_u^2 g(u)=\Delta_u(\Delta_u g(u))=\left(4d(d+2)\partial_u^2+16(d+2)u\partial_u^3+16u^2\partial_u^4\right)g(u).\nonumber
\EEQ

For illustration, we solve the equation (\ref{eq:galilei}) for $n=2$, that is $z=4$. In this case, one has
 \BEQ
 \left((d+2)\partial_p+2(d+8)p\partial_p^2+8p^2\partial^3_p+\frac{\mu}{8}\right)f(p)=0. \label{spacial}
 \EEQ
There are three independent solutions which are sought as Frob\'enius series
 \BEQ
 f(p)=\sum_{k=0}^\infty a_k p^{k+\sigma}\;\;,\;\; a_0\ne 0.\label{eq:f2sol}
 \EEQ
where the exponent $\sigma$ is found from the characteristic equation  $\sigma\left(8\sigma^2+2\sigma(d-4)+2-d\right)=0$
with solutions $\sigma=0, \demi, \demi-\frac{d}{4}$. Standard methods give a recursion relation for the coefficients $a_m$ and lead to,
for $n=2$ or equivalently $z=4$
  \BEA
  f(p) & = & f_0\:{}_0F_2\left(\demi, \demi +{d\over 4}; -\frac{\mu p}{64}\right)+f_1\,p^{1/2}\:{}_0F_2\left(\frac{3}{2}, {d\over 4}+1;
  -\frac{\mu p}{64}\right)\nonumber\\
  &  & + f_2\,p^{1/2-d/4}\:{}_0F_2\left(1-{d\over 4},\frac{3}{2}-{d\over 4}; -\frac{\mu p}{64}\right).\label{2pointz4}
  \EEA
where ${}_0F_2$ is a generalised hyper-geometric function and $f_{0,1,2}$ are normalisation constants.
One could use the explicitly known asymptotic forms of ${}_0F_2(a,b;x)$ for $x\to\pm\infty$
in order to identify the sub-space of
solutions which do not diverge for $|\mu p|\gg 1$ \cite{Wright35,Henkel11},
but we shall not need this explicitly later.

\subsection{Generic dynamical exponent}

The covariance under the generalised Galilei transformations of the representation (\ref{eq:z2mplus1algebra}) requires the
$d$ conditions
\BEQ \label{detrzgen}
\left(\tau\partial_{r_j}\nabla^{z-2}_{\vec{r}}+\mu r_j\right)f(\tau,\vec{r})=0 \;\; , \;\; j=1,...,d.
\EEQ
The Frob\'enius series method from above cannot be used, because of the fractional derivative. However, eq.~(\ref{detrzgen})
can be solved in Fourier space where
\BEQ
f(\tau,\vec{r})=\frac{1}{(2\pi)^{d}}\int_{\R^d}\!\D\vec{k}\; e^{\II \vec{k}\cdot\vec{r}}{\wht f}(\tau,\vec{k}) \;.
\EEQ
This leads to the correspondences $\partial_j \mapsto \II k_j$,
$r_j \mapsto \II \frac{\partial}{\partial k_j}$ and
$\nabla_{\vec{r}}^{\alpha} \mapsto (\II|\vec{k}|)^{\alpha}$.
It follows that equation~(\ref{detrzgen}) takes the following form in Fourier space:
\BEQ
(\mu\partial_{k_j}+\II^{z-2}\tau k_j|\vec{k}|^{z-2}){\wht f}(\tau, \vec{k})=0, \quad j=1,...,d\label{furiergal}
\EEQ
and has the the solution
\BEQ
{\wht f}(\tau, \vec{k})=f_0(\tau) \exp\left[-\frac{\II^{z-2}}{z}\frac{\tau}{\mu} |\vec{k}|^{z}\right].\label{solutionfurier}
\EEQ
This is rewritten in the direct space as follows
\BEA
f(\tau, \vec{r}) &=& \frac{f_0(\tau)}{(2\pi)^{d}} \int_{\R^d}\!\D\vec{k}\:
\exp\left[\II\vec{k}\cdot\vec{r}-\frac{\II^{z-2}}{z}\frac{\tau}{\mu} |\vec{k}|^{z}\right]
= \frac{f_0(\tau)}{(2\pi)^d} I_{\beta}(\vec{r}) \;\; ; \;\; \beta := \frac{\II^{z-2}\tau}{z \mu}
\label{integralsolution}
\EEA
with the following abbreviation
\BEQ
I_{\beta}(\vec{r}) :=\int_{\R^d}\!\D\vec{k}\:
\exp\left[\II\vec{k}\cdot\vec{r}-\beta|\vec{k}|^{z}\right]
\label{keyintegral}
\EEQ
In what follows, we shall admit $\beta\in\C $, but suitably restricted such that integral converges
(i.e. $\Re \beta>0$ would be sufficient).
To evaluate the integral (\ref{keyintegral}) formally,
introduce $d$-dimensional in spherical coordinates
\BEA
k_1 &=& k\cos\theta_1\nonumber\\
k_2 &=& k\sin\theta_1\cos\theta_2\nonumber\\
\vdots \\
k_{d-1} &=& k\sin\theta_1\sin\theta_2...\sin(\theta_{d-2})\cos(\theta_{d-1})\nonumber\\
k_{d} &=& k\sin\theta_1\sin\theta_2...\sin(\theta_{d-2})\sin(\theta_{d-1}),
\nonumber\label{spcord}
\EEA
where $k=|\vec{k}|$, $\theta_j \in[0,\pi)$ for $j=1,\ldots d-2$ and $\theta_{d-1}\in[0,2\pi)$. The jacobian is
\BEQ
\frac{\partial(k_1,k_2,\ldots,k_d)}{\partial(k,\theta_1,\ldots,\theta_{d-1})}
=k^{d-1}\sin^{d-2}\theta_1\sin^{d-3}\theta_2
\ldots \sin(\theta_{d-2})\nonumber
\EEQ
Then the integral (\ref{keyintegral}) becomes
\BEA
I_{\beta}(\vec{r}) & = & \int_0^{\infty}\!\D k\: k^{d-1}e^{-\beta k^z}
\int_0^{\pi}\!\D\theta_1 e^{\II  kr\cos\theta_1}\sin^{d-2}\theta_1\nonumber\\
& &\times \int_0^{\pi}\!\D \theta_2 \sin^{(d-1)-2}\theta_2 \ldots
\int_0^{\pi}\!\D \theta_{d-2} \sin^{(d-1)-(d-2)}\theta_{d-2}\int_0^{2\pi}\!\D \theta_{d-1}\nonumber\\
&=& S_{d-1}\int_0^{\infty}\!\D k\: k^{d-1}\,e^{-\beta k^z}\int_0^{\pi}\!\D \theta_1\, e^{\II kr\cos\theta_1}  \sin^{d-2}\theta_1
\nonumber \\
&=&(2\pi)^{d/2}r^{-(d-2)/2}\int_0^{\infty}\!\D k\: k^{d/2}\, J_{d/2-1}(kr)\, e^{-\beta k^z}
\nonumber \\
&=& (2\pi)^{d/2} r^{-d} \int_0^{\infty} \!\D u\: u^{d/2}\, J_{d/2-1}(u) \,e^{-(\beta r^{-z}) u^z}
\label{betweenr}
\EEA
where we used \cite[eq.~(8.411.7)]{GR}
in the third line in order to express the last angular integral in terms of a Bessel function
$J_{\nu}(w)$ and $S_{d}=2 \pi^{d/2}/\Gamma(d/2)$ denotes the surface of the $d$-dimensional sphere.

In order to display explicitly the space-time scaling of the scaling function $f(\tau,\vec{r})$,
we re-express $\beta = \alpha \tau$ such that $\alpha=\II^{z-2}/(z\mu)$ and then have
\BEQ
f(\tau,\vec{r}) = \frac{f_0(\tau)}{(2\pi)^{d/2}} r^{-d} \Lambda\left( \frac{\alpha\tau}{r^z}\right)
\EEQ
where the scaling function $\Lambda$ is given by the above integral.
Applying scale-invariance, gives an obvious
differential equation for the last remaining function, hence $f_0(\tau)=f_{(0)} \tau^{d/z}$.
Finally, use the expansion of the Bessel function $J_{\nu}(w)=\sum_{n=0}^{\infty}\frac{(-1)^n(w/2)^{\nu+2n}}{n!\Gamma(\nu+n+1)}$,
in order to convert the scaling function $f(\tau,\vec{r}) = (2\pi)^{-d/2} f_{(0)} (r^z/\tau)^{d/z} \Lambda(\alpha\tau r^{-z})$
into a series. This gives the final result
\BEQ \label{scalingfunctionfinal}
f(\tau,\vec{r}) = f_{00}\, \frac{\Gamma(d/2)}{\Gamma(d/z)}
\sum_{n=0}^{\infty} (-1)^n \frac{\Gamma\left(\frac{2n+d}{z}\right)}{n! \Gamma\left(n+\frac{d}{2}\right)}
\left( \frac{r^2}{4 (\alpha \tau)^{2/z}}\right)^{n}
\EEQ
and where $f_{00}$ is a normalisation constant. Clearly,
this series has an infinite radius of convergence for $z>1$.

The sought form of the two-point function $F$ is given by
eqs.~(\ref{4.8},\ref{scalingfunctionfinal}). The special case of even values
of $z$ treated above is compatible with this form.\footnote{For example, in the case $z=4$,
one may separate the sum into even and
odd terms. Repeated application of the duplication formula
$\Gamma(2k)=\pi^{-\demi}2^{2k-1}\Gamma(k)\Gamma(k+1/2)$ shows that
the sums arising are exactly of the kind found in (\ref{2pointz4}).
We have also seen that the form (\ref{scalingfunctionfinal})
has a Fourier transformation, which is not obvious for all solutions in (\ref{2pointz4}).}

\section{Spherical model and field-theoretical description}

\subsection{Spherical model with conserved order parameter}

The spherical model was conceived in 1953 by
Berlin and Kac as an exactly solvable mathematical model for strongly interacting spins
and has proved ever since to be an useful model for trying out more general ideas.
It is usually defined in terms of real spin variable $S(t, \vec{x})$
attached to each site $\vec{x}$ of the hyper-cubic lattice $\Lambda\subset\mathbb{Z}^d$
and depending on time $t$, subject to the mean spherical constraint
\BEQ
\left\langle\sum_{\vec{x}\in\Lambda}S(t,\vec{x})^2\right\rangle={\cal N},\label{constraint}
\EEQ
where $\cal N$ is the number of sites. The spin Hamiltonian is ${\cal H}=-\sum_{(\vec{x},\vec{y})}S_{\vec{x}}S_{\vec{y}}$ where
the sum is over pairs of nearest neighbours. At equilibrium, this leads
to a second-order phase transition with a critical temperature $T_c>0$ for all spatial dimensions $d>2$
and where the critical exponents have non-mean-field values for $d<4$,
see e.g. \cite{Joyce72} and references therein.
We consider here a kinetics given by a Langevin equation with a conserved order parameter
(model B in the terminology of \cite{Hohenberg77})
\BEQ
\partial_tS(t,\vec{x})=-\nabla^2_{\vec{x}}[\nabla^2_{\vec{x}}S(t,\vec{x})+\mathfrak{z}(t)S(t,\vec{x})+h(t,\vec{x})]+\eta(t,\vec{x})
\label{blangevin},
\EEQ
where $\mathfrak{z}(t)$ is the Lagrange multiplier fixed by the mean spherical constraint
and the coupling to the heat bath with the critical temperature $T_c$ is described by a Gaussian noise
$\eta$ of vanishing average and variance
\BEQ
\langle\eta(t,\vec{x})\eta(t',\vec{x}')\rangle = -2T_c\nabla^2_{\vec{x}}\delta(t-t')\delta(\vec{x}-\vec{x}')\label{conservednoise}
\EEQ
and $h(t,\vec{x})$ is a small external magnetic field (required for the computation of the response)
\cite{Kissner92,Majumdar95,Sire04,Baumann07}. This constitutes a
qualitatively reasonable model for the kinetics of phase-separation, for instance in alloys.

Essentially the same equations can also be used to
describe very different physical situations. For example, the growth
of interfaces on a substrate with a conservation of
particles along the interface is described by the Mullins-Herring equation
\cite{Mullins63}
(or the equivalent Wolf-Villain model \cite{Wolf90})
which in our notation is simply given by the Langevin equation (\ref{blangevin})
with $\mathfrak{z}=0$. Several inequivalent variants exist: one may either take the conserved noise (\ref{conservednoise}) or
else consider the non-conserved centred noise with variance
$\langle\eta(t,\vec{x})\eta(t',\vec{x}')\rangle = 2T_c\delta(t-t')\delta(\vec{x}-\vec{x}')$,
see \cite{Roethlein06} for details.

Here, we shall concentrate on the Mullins-Herring model with the conserved noise (\ref{conservednoise}).\\

Since correlators and response have already been studied in detail in the literature \cite{Kissner92,Majumdar95,Godreche04,Sire04,Baumann07}
it is enough to quote the results.
For our purposes, we merely need the full space-time response in the conserved spherical model
for $d>4$ or equivalently the Mullins-Herring equation for any $d$, which reads
 \BEA
  && R(t,s;\vec{r})= \frac{\sqrt{\pi}}{2^{3d/2}\pi^{d/2}\Gamma(d/4)}
  (t-s)^{-(d+2)/4}\left[{}_0F_2\left(\demi,\frac{d}{4};\frac{r^4}{256(t-s)}\right)\right.\nonumber\\
  && -\left.\frac{8}{d}\frac{\Gamma(\frac{d}{4}+1)}{\Gamma(\frac{d}{4}+\demi)}\left(\frac{r^2}{16\sqrt{t-s}}\right)
  {}_0F_2\left(\frac{3}{2},\frac{d}{4}+\demi;\frac{r^4}{256(t-s)}\right)\right].\label{exactresponse}
  \EEA
We want to compare this  with the
$\mathfrak{age}(d)$-covariant two-point function, as obtained in the previous section from the
non-local representation (\ref{eq:z2nalgebra}), with $z=4$.
Before, we can do this, however, we still have to show how to relate
a dynamical symmetry of a deterministic equation such as (\ref{fracsch}) with the properties of a solution of a stochastic Langevin
equation (\ref{blangevin}).

\subsection{Field-theoretical description}

In order to compare co-variant response computed from the non-local representations and derived in section~3,
we appeal to standard methods in non-equilibrium field-theory, see e.g. \cite{Tauber05, Tauber07} for introductions.
Thereby, we shall generalise previous applications of the method \cite{Picone04,Baumann05b,Baumann07,Baumann07b,Durang09,Stoimenov09}
to the present case of non-local representations. Consider the Langevin equation
 \BEQ
 \partial_t\phi=-\frac{1}{4\mu}\nabla^2_{\vec{r}}
 \left(-\nabla^2_{\vec{r}}\phi+v(t)\phi\right)+\eta\label{mblangevin}
 \EEQ
with the conserved centred gaussian noise
\BEQ
\langle\eta(t,\vec{r})\eta(t',\vec{r}')\rangle
= -\frac{T_c}{2\mu}\nabla^2_{\vec{r}}\delta(t-t')\delta(\vec{r}-\vec{r}').\label{mconservednoise}
\EEQ
Following \cite{Tauber05, Tauber07, Baumann07}, the Janssen-de Dominicis action is
\BEA
{\cal J}(\phi,\wit{\phi}) & = & \int \!\D u \D\vec{R}\:
\left[\tilde\phi\left(\partial_u-\frac{1}{4\mu}\nabla^2_{\vec{R}}(\nabla^2_{\vec{R}}-v(u))\right)\phi\right]
\nonumber\\
& & +\frac{T}{4\mu}\int \!\D u \D\vec{R}\:
\wit{\phi}(u,\vec{R})(\nabla^2\wit{\phi}(u,\vec{R}))+{\cal J}_{init}(\phi,\wit{\phi})\label{jaction}
\EEA
where
\BEQ
{\cal J}_{init}(\phi,\wit{\phi})=\demi\int \!\D\vec{R} \D\vec{R}'\:
\wit{\phi}(0,\vec{R})\left\langle\phi(0,\vec{R})\phi(0,\vec{R}'\right\rangle\wit{\phi}(0,\vec{R}')\label{noiseinit}
\EEQ
with complete analogy with non-conserved case \cite{Mazenko04,Picone04}.

Averages of an observable ${\cal A}$ are defined as usual by the functional integral
\BEQ
 \langle{\cal A}\rangle=\int{\cal D}[\phi]{\cal D}[\tilde\phi]\: {\cal A}[\phi] \exp(-{\cal J}(\phi,\wit{\phi})).\label{meanvariable}
 \EEQ\\
Next, one decomposes the action into a deterministic part ${\cal J}_0$ and a noise part ${\cal J}_b$
\BEQ
 {\cal J}[\wit{\phi},\phi])={\cal J}_0[\wit{\phi},\phi])+{\cal J}_b[\wit{\phi}]
 \EEQ
with
\BEA
 && {\cal J}_0(\phi,\wit{\phi})= \int \!\D u \D\vec{R}\:
 \left[\tilde\phi\left(\partial_u-\frac{1}{4\mu}\nabla^2_{\vec{R}}(\nabla^2_{\vec{R}}-v(u))\right)\phi\right]\label{detaction}\\
&& {\cal J}_b={\cal J}_{init}+{\cal J}_{th};\quad {\cal J}_{th}=\frac{T}{4\mu}\int \!\D u \D\vec{R}\:
\wit{\phi}(u,\vec{R})(\nabla^2\wit{\phi}(u,\vec{R})).\label{noiseaction}
\EEA
The point of this decomposition is that the action ${\cal J}_0$
is invariant under the non-local representations constructed in section~2.
We call the theory with action ${\cal J}_0$ {\em noise free}
or deterministic and denote the average with respect to it by
$\langle...\rangle_0$.
Then any average of the full theory can then be computed by formally expending around the noise-free theory
\BEQ
\langle{\cal A}\rangle=\langle{\cal A}\exp(-{\cal J}_b)\rangle_0.\label{average}
\EEQ

We can focus on the deterministic theory with $v=0$ because of the following observation:
{\em If $\phi(t,\vec{r})$ is a solution of equation (\ref{mblangevin}) with $\eta=0$ then the field}
\BEQ
\psi(t,\vec{r}) :=\exp\left(\frac{1}{4\mu}\int_0^t \!\D\tau\: v(\tau)\nabla^2_{\vec{r}}\right)\phi(t,\vec{r})\label{gaugetransform}
\EEQ
{\em fulfils the same equation (\ref{mblangevin}) with $v=0$ (and $\eta=0)$.}
Hence, it is enough to consider first the problem with $v=0$
and apply the inverse of the transformation (\ref{gaugetransform})
for treating the case $v\ne 0$, in order to implement the breaking of time-translation invariance.
On the other hand, the case $v=0$ is relevant for $d>4$
in the spherical model and for any $d$ in Mullins-Herring model, which are the cases we are interested in.

The noise-free theory is given by the equation (\ref{mblangevin}) with
$\eta=0, v=0$ or by the action ${\cal J}_0$ with $v=0$.
This equation co\"{\i}ncides with the equation (\ref{schrodinger})
in section 2 with $n=2$ when $x+\xi-(d+2)/2=0$.
If the fields $\phi,\wit{\phi}$ are indeed quasi-primary,
that is they transform under non-local representation (\ref{eq:z2nalgebra})
of the ageing algebra $\mathfrak{age}(d)$, it follows that
 \BEQ
  \phi=\phi_{(\mu,x,\xi)}\quad,\quad\wit{\phi}=\wit{\phi}_{(-\mu,\tilde{x},\tilde{\xi})}\label{scalingdimensions}
  \EEQ
are characterised with opposite``mass" parameter and scaling dimensions $(x,\xi),(\tilde{x},\tilde{\xi})$ correspondingly, with relation
\BEQ
x+\xi=\tilde{x}+\tilde{\xi}=(d+2)/2.\label{relationscalingdimensions}
\EEQ
The noise-free theory then is invariant under the non-local representation (\ref{eq:z2nalgebra}) of $\mathfrak{age}(d)$.
The Bargman super-selection rule \cite{Bargman54} holds in term of the ``mass" parameter $\mu$ that is
\BEQ
\langle\phi_1(t_1,\vec{r}_1)...\phi_n(t_{n},\vec{r}_{n})\rangle_0=0\label{mbargman}
\EEQ
unless $\mu_1+\ldots\mu_n=0$. On the other hand because of the Gaussian structure of the action ${\cal J}_0$
\cite{Baumann07} of the noise-free theory, Wick's theorem applies and gives \cite{Zinn02}
\BEA
&& \langle\phi_1(t_1,\vec{r}_1)...\phi_{2n}(t_{2n},\vec{r}_{2n})\rangle_0
=\sum_{\mbox{\rm\footnotesize All possible pairings P of $1,2,...2n$}}\nonumber\\
&& \langle\phi_{P_1}(t_{P_1},\vec{r}_{P_1})\phi_{P_2}(t_{P_2},\vec{r}_{P_2})\rangle_0
...\langle\phi_{P_{2n-1}}(t_{P_{2n-1}},\vec{r}_{P_{2n-1}})\phi_{P_{2n}}
(t_{P_{2n}},\vec{r}_{P_{2n}})\rangle_0.\label{pairing}
\EEA
As in the non-conserved case \cite{Picone04,Baumann07},
it follows that the response function is independent of noise,
while in noise-free theory it is given by
\BEQ
R(t,s;\vec{x}-\vec{y}) :=\langle\phi(t,\vec{x})\nabla^2_{\vec{y}}\wit{\phi}(s,\vec{y})\rangle_0
=\nabla^2_{\vec{r}}F^{(2)}(t,s;\vec{r})\label{response}
\EEQ
where $F^{(2)}(t,s;\vec{r})$ is the two-point function computed in the previous section in eqs.~(\ref{4.8},\ref{2pointz4});
after taking time-translation invariance into account, which we achieve under important assumption
$x+\tilde{x}=d$, in addition to the relations (\ref{relationscalingdimensions}). It follows
\BEA
 R(t,s;\vec{r}) & = & (t-s)^{-d/4}\nabla^2_{\vec{r}}f\left(\frac{\vec{r}^4}{t-s}\right)=(t-s)^{-(d+2)/4}\Delta_pf(p)\nonumber\\
& = & 4(t-s)^{-(d+2)/4}((d+2)p^{\demi}\partial_p+4p^{\frac{3}{2}}\partial^2_p)\times\nonumber\\
 & & \times  \left[f_0\;\:{}_0F_2\left(\demi, \demi +{d\over 4}; -\frac{\mu p}{64}\right)
 +f_1p^{1/2}\:{}_0F_2\left(\frac{3}{2}, {d\over 4}+1; -\frac{\mu p}{64}\right)\right.\nonumber\\
 &  &+ \left. f_2p^{1/2-d/4}\:{}_0F_2\left(1-{d\over 4},\frac{3}{2}-{d\over 4}; -\frac{\mu p}{64}\right)\right]
 \nonumber\\
 & = & (t-s)^{-(d+2)/4}\left[f'_1\;\:{}_0F_2\left(\demi,{d\over 4}; -\frac{\mu p}{64}\right)
 +f'_0p^{1/2}\:{}_0F_2\left(\frac{3}{2}, {d\over 4}+\demi; -\frac{\mu p}{64}\right)\right.\nonumber\\
 &  &+ \left. f'_2p^{1-d/4}\:{}_0F_2\left(\frac{3}{2}-{d\over 4},2-{d\over 4}; -\frac{\mu p}{64}\right)\right].
\label{finalresponse}
\EEA
While the response function must be regular for $\vec{r}=0$ and goes to $0$
when $\vec{r}\to \infty$ the third term is eliminated and the
constants $f'_0$ and $f'_1$ must be related.
Using the known long-term behaviour of the hyper-geometric function \cite{Wright35,Henkel11}
we recover exactly the result (\ref{exactresponse}).

Therefore, we have seen that the exact result (\ref{exactresponse}),
rather than being interpreted as in \cite{Baumann07}
a consequence of a local symmetry,
the closure of which as a Lie algebra is somewhat artificial and {\it ad hoc}, may also be seen
as evidence of a non-local representation (\ref{eq:z2nalgebra})
of the well-known Lie algebra $\mathfrak{age}(d)$, where the closure is naturally
provided by the same equation of motion which is under study~!

\section{Conclusions}

In trying to construct a closed set of Lie algebra generators for
generalised scale-transformations with an arbitrary dynamical
exponent $z$, we have been led to consider non-local representations
of the the ageing algebra $\mathfrak{age}(d)$. Clearly, the development of such algebraic techniques
can only be a first step to hopefully address in the future
the considerably more difficult question of genuinely non-linear equations.
In this spirit, the present work is intended as a case study and is meant to provide at least one explicit example.
Our construction was guided by the previously known results for $d=1$ space dimensions \cite{Henkel11},
herein extended to generic $d>1$. It turns out that the result depends on the value of the dynamical exponent $z$.

Conceptually, we have slightly extended the usual definition
of the notion of {\em dynamical symmetry}, see for example \cite{Niederer72}.
Conventionally, the infinitesimal generator $X$ of a
dynamical symmetry of the equation of motion ${\cal S}\psi=0$ must satisfy
$[{\cal S},X] = \lambda_X {\cal S}$ as an operator. Furthermore, it is conventionally admitted that
$\lambda_X$ should be a scalar or a function. Here, we have allowed for the possibility that it may also
become an {\em operator}.
In our construction of non-local representations of ageing algebra this appears only
for the generator of special (conformal) transformations $X_1$,
for which $\lambda_{X_1}$ is a {\em differential operator}.

In addition we use a {\em quotient} space construction we use to close the algebra.
Such a construction appears to be necessary whenever
the dynamical exponent $z\ne 1,2$.
At least, this appears to be a more elegant and hopefully mathematically more
sober way to achieve the closure
of the various generators into a Lie algebra,
and appears to us preferable to the more traditional approach where several
{\it ad hoc} super-selection rules had to be imposed in order
to cut off the infinite set of non-closing operators, see
\cite{Henkel02,Baumann07,Henkel07,Henkel10}.

Several details depend on the value of $z$
\begin{enumerate}
\item For $z$ even, the algebra (\ref{eq:z2nalgebra}) contains the usual {\em local} generators of space-translations,
spatial rotations, dynamical scaling and global phase shift. In addition, the further $d+1$ generators of
generalised Galilei-transformation and special transformations are {\em non-local} and are
constructed with linear differential operators of order $z-1$.
By analogy with  the $1D$ case \cite{Henkel11},
we suspect that these might be interpreted as generating transformation
of distribution functions of the positions, rather than
{\it bona fide} coordinate transformations.
The constructed representation acts naturally as a dynamical symmetry on the
solution space of the simple linear Schr\"odinger equation we considered,
and should be viewed essentially as a toy model.
The example studied here might be the first step towards an understanding how to use
such non-local transformations in applications to the
non-equilibrium physics of strongly interacting particles.

We recall that in the context of interface growth with conserved dynamics,
exactly the kind of non-local generalised  Galilei-transformation
we have studied here has already been introduced in analysing the stochastic equation
(related to molecular beam epitaxy ({\sc mbe}))
\BEQ
\partial_t \phi = - \nabla^2 \left[ \nu \nabla^2 \phi + \frac{\lambda}{2} (\nabla \phi)^2 \right] + \eta
\EEQ
where $\eta$ is a centred and conserved gaussian noise and $\nu,\lambda$ are constants \cite{Sun89}.
Indeed, they show that Galilei-invariance leads to a non-trivial hyperscaling relation,
expected to be exact and checked in the lowest orders of the loop expansion. In particular,
they obtain $z=4$ in $d=2$ space dimensions \cite{Sun89}.
Similar results exist for slightly different non-linear conserved growth
equations \cite{Sarma94}.
We hope to return to a symmetry analysis of these non-linear equations in the future. In any case, the
available evidence that generalised Galilei-invariance could survive the loop expansion is very encouraging.

Since the non-local representations only close over solution of the same equation,
we deal with an {\it on shell} representation.\footnote{Its sub-algebra,
however, which includes the dynamical scaling and Galilei algebra is
{\it off shell} that is, it is closed without any conditions.}
It would be interesting to see whether a systematic relationship
with {\em conditional symmetries} in partial differential equations can be found,
see \cite{Cherniha10} and references therein.

\item For generic values of $z\geq 2$, the generalisation from the one-dimensional
case requires the explicit introduction of some kind
of fractional derivative. For our purposes, the Riesz fractional derivative turned out
to have the required algebraic properties.
In addition, the result derived for the co-variant two-point function is
compatible with the directly treatable case when $z$ is even.
We are not aware of confirmed physical applications with generic values of $z$.
It remains to be seen if similar symmetries of other non-local equations can be found.

On the other hand, the generalisation proposed here in terms of strongly non-local Riesz fractional derivatives
into the generators permits at least
to close the representation on the solution space of a single partial differential equation.
Another advantage of this realisation is that it
naturally includes the case $z$ even,
when the fractional derivatives become the powers of the Laplacian.
It formally can be interpreted as
dynamical symmetry algebra and may have unexpected applications which we hope to find in the future.

\item One important difference between $z=2n$ even and $z$ generic is that in the latter case, the only technique available of
solving fractional differential equations, namely by Fourier transformation,
gives merely a subset of the solutions which can be found by Frob\'enius-type
series for $z=2n$.

It remains an open question whether different choices of the fractional derivative are possible, which may also
imply that different techniques for finding $n$-point functions will have to be used.
Our aim has been to provide at least one worked-out example and we do not claim that the technical
choices made here should be unique.

It also remains to be seen how to interpret the `finite' transformations obtained from the exponentiated
non-local generators $\exp(\eps X)$ geometrically.
\item For $1<z<2$, the radius of convergence of the series representations of the two-point function is still infinite and for
$z=1$ it is finite. It could be of interest to see whether the specific form (\ref{scalingfunctionfinal})
might already be of relevance in concrete models.
Alternatively, it might be
conceivable that non-local representations, constructed from the standard {\em local} representations of the
conformal Galilei algebra ({\sc cga}) which  have $z=1$ \cite{Martelli09,Cherniha10} could cover the range $1\leq z<2$. This
remains an open question for future work.
\end{enumerate}
Clearly, the extension of the present method to {\em non-linear} equations remains another important open question.

\noindent
{\bf Acknowledgements:} Most of the work on this paper was done during the visit of S.S. at
the Universit\'e Henri Poincar\'e Nancy I (now Universit\'e de Lorraine Nancy).
The visit was supported in parts: by the Bulgarian NSF grant {\it DO 02-257}
and the Universit\'e Henri Poincar\'e Nancy I.
M.H. was partly supported by the Coll\`ege Doctoral Nancy-Leipzig-Coventry
(Syst\`emes complexes \`a l'\'equilibre et hors \'equilibre) of UFA-DFH.


\begin{thebibliography}{999}
\bibitem{Bala08} K. Balasubramanian and J. McGrevy, Phys. Rev. Lett. {\bf 101}, 061601 (2008).
\bibitem{Bargman54} V. Bargman, Ann. of Math. {\bf 56}, 1 (1954).
\bibitem{Baumann05b} F. Baumann, S. Stoimenov and M. Henkel, J. Phys. {\bf A39}, 4095 (2006).
\bibitem{Baumann07} F. Baumann and M. Henkel, J. Stat. Mech. P01012 (2007).
\bibitem{Baumann07b} F. Baumann, S.B. Dutta, and M. Henkel, J. Phys. A: Math. Gen. {\bf 40}, 7389 (2007).
\bibitem{Bray94a} A.J. Bray, Adv. Phys. {\bf 43}, 357 (1994).
\bibitem{Cherniha10} R. Cherniha and M. Henkel, J. Math. Anal. Appl. {\bf 369}, 120 (2010).
\bibitem{Cugl02} L.F. Cugliandolo, in {\it Slow Relaxation and
non equilibrium dynamics in condensed matter}, Les Houches Session 77 July 2002,
J-L Barrat, J Dalibard, J Kurchan, M V Feigel'man eds, Springer(Heidelberg 2003) ({\tt cond-mat/0210312}).
\bibitem{Durang09} X. Durang and M. Henkel, J. Phys. A: Math. Theor. {\bf 42}, 395004 (2009).
\bibitem{Duval09} C. Duval and P.A. Horv\'athy, J. Phys. A: Math. Theor. {\bf 42}, 465206 (2009).
\bibitem{Fuertes09} C.A. Fuertes and S. Moroz, Phys. Rev. {\bf D79}, 106004 (2009).
\bibitem{Godreche04} C. Godr\`eche, F. Krzakala and F. Ricci-Tersenghi, J. Stat. Mech. P04007 (2004).
\bibitem{GR} I.S. Gradshteyn and I.M. Ryzhik,
{\it Table of integrals, series and products}, 4$^{\rm th}$ ed., Academic Press (London 1980)
\bibitem{Gray13} N. Gray, D. Minic and M. Pleimling, {\tt arxiv:1301.6368}
\bibitem{Havas78} P. Havas and J. Plebanski, J. Math. Phys. {\bf 19}, 482 (1978).
\bibitem{Henkel94} M. Henkel, J. Stat. Phys. {\bf 75}, 1023 (1994).
\bibitem{Henkel97} M. Henkel, Phys. Rev. Lett. {\bf 78}, 1940 (1997).
\bibitem{Henkel02} M. Henkel, Nucl. Phys. {\bf B641}, 405 (2002).
\bibitem{Henkel03a} M. Henkel and J. Unterberger, Nucl. Phys. {\bf B660}, 407 (2003).
\bibitem{Henkel06a} M. Henkel, T. Enss and M. Pleimling, J. Phys. A Math. Gen. {\bf 39}, L589 (2006).
\bibitem{Henkel06b} M. Henkel, R. Schott, S. Stoimenov and J. Unterberger,
Confluentes Mathematici {\bf 4}, 1250006 (2012).
\bibitem{Henkel07} M. Henkel and F. Baumann, J. Stat. Mech. P07015 (2007).
\bibitem{Henkel10} M. Henkel and M. Pleimling,
{\it Non-equilibrium phase transitions vol. 2: ageing and dynamical scaling far
from equilibrium}, Springer (Heidelberg 2010).
\bibitem{Henkel11} M. Henkel and S. Stoimenov, Nucl. Phys {\bf B847}, 612 (2011).
\bibitem{Henkel12} M. Henkel, {\it Causality from dynamical symmetry: an example from local scale-invariance},
to appear in the proceedings of AGMP-7 (Mulhouse, France, oct 2011), {\tt arxiv:1205.5901}
\bibitem{Henkel12a} M. Henkel, J.D. Noh and M. Pleimling, Phys. Rev. {\bf E85}, 030201(R) (2012).
\bibitem{Henkel13a} M. Henkel, Nucl. Phys. {\bf B869}, 212 (2013).
\bibitem{Henkel13b} M. Henkel and S. Rouhani, {\tt arxiv:1302.7136}
\bibitem{Hohenberg77} P. Hohenberg and B.I. Halperin, Rev. Mod. Phys.  {\bf 49}, 435 (1977).
\bibitem{Ivash97} E.V. Ivashkevich, J. Phys. A: Math. Gen. {\bf 30}, L525 (1997).
\bibitem{Janssen92} H.-K. Janssen, in G. Gy\"orgi {\it et al.} (eds), {\it From phase-transitions to chaos},
World Scientific (Singapour 1992).
\bibitem{Joyce72} G.S. Joyce, in C. Domb and M. Green (eds),
{\it Phase transitions and critical phenomena}, vol. 2,
Academic Press (London 1972); p. 375
\bibitem{Kardar86} M. Kardar, G. Parisi and Y.-C. Zhang, Phys. Rev. Lett. {\bf 56}, 889 (1986).
\bibitem{Kissner92} J.G. Kissner, Phys. Rev. {\bf B46}, 2676 (1992).
\bibitem{Krug91} J. Krug and P. Meakin, Phys. Rev. Lett. {\bf 66}, 703 (1991).
\bibitem{Lukierski06} J. Lukierski, P.C. Stichel and W.J. Zakrewski, Phys. Lett. {\bf A357}, 1 (2006);
Phys. Lett. {\bf B650}, 203 (2007).
\bibitem{Mainardi07} F. Mainardi and R. Gorenflo, Fractional Calculus and Applied Analysis, {\bf 10}, 269 (2007).
\bibitem{Majumdar95} S.N. Majumdar and D.A. Huse, Phys. Rev. {\bf E52}, 270 (1995)
\bibitem{Martelli09} D. Martelli and Y. Tachikawa, JHEP 1005:091 (2010). [{\tt arxiv:0903.5184}]
\bibitem{Mazenko04} G.F. Mazenko, Phys. Rev. {\bf E69}, 016114 (2004).
\bibitem{Miller93} K.S. Miller and B. Ross,
{\it An introduction to the fractional calculus and fractional
differential equations}, Wiley (New York 1993).
\bibitem{Minic08} D. Minic and M. Pleimling, Phys. Rev. {\bf E78}, 061108 (2008).
\bibitem{Minic12} D. Minic, D. Vaman and C. Wu, Phys. Rev. Lett. {\bf 109}, 131601 (2012)
\bibitem{Mullins63} W.W. Mullins, in N.A. Gjostein and W.D. Robertson (eds),
{\it Metal surfaces: Structure, energetics,
kinetics}, American Society of Metals (Metals Park (Ohio) 1963).
\bibitem{Negro97} J. Negro, M.A. del Olmo and A. Rodr\'{\i}guez-Marco,
J. Math. Phys. {\bf 38}, 3786 and 3810 (1997).
\bibitem{Niederer72} U. Niederer, Helv. Phys. Acta {\bf 45}, 802 (1972).
\bibitem{Picone04} A. Picone and M. Henkel, Nucl. Phys. {\bf B688} 217 (2004).
\bibitem{Podlubny99} I. Podlubny, {\it Fractional differential equations}, Academic Press (New York 1999).
\bibitem{Popkov11} V. Popkov and G.M. Sch\"utz, J. Stat. Phys. {\bf 142}, 627 (2011).
\bibitem{Roethlein06} A. R{\"o}thlein, F. Baumann and M. Pleimling,  Phys. Rev. {\bf E74}, 061604 (2006);
Erratum {\bf E76}, 019901 (2007).
\bibitem{Sarma94} S. Das Sarma and R. Kotlyar, Phys. Rev. {\bf E50}, R4275 (1994).
\bibitem{Samko93} S.G. Samko, A.A. Kilbas and O.I. Marichev,
{\it Fractional integrals and derivatives: theory and applications}, Gordon and Breach (Amsterdam 1993).
\bibitem{Sire04} C. Sire, Phys. Rev. Lett. {\bf 93}, 130602 (2004).
\bibitem{Son08} D.T. Son, Phys. Rev. {\bf D78}, 106005 (2008).
\bibitem{Spohn99} H. Spohn, Phys. Rev. {\bf E60}, 6411 (1999).
\bibitem{Stoimenov05} S. Stoimenov and M. Henkel, Nucl. Phys. {\bf B723}, 205 (2005).
\bibitem{Stoimenov09} S. Stoimenov, Fortschr. Phys. {\bf 57}, 711 (2005).
\bibitem{Stru78} L.C.E. Struik,
{\it Physical Ageing in Amorphous Polymers and Other Materials}, Elsevier (Amsterdam 1978)
\bibitem{Sun89} T. Sun, H. Guo and M. Grant, Phys. Rev. {\bf A40}, 6763 (1989).
\bibitem{Tauber05} U.C. T\"auber, M. Howard and B.P. Vollmayr-Lee, J. Phys. {\bf A38}, R79 (2005).
\bibitem{Tauber07} U.C. T\"auber, in M. Henkel, M. Pleimling and R. Sanctuary (eds)
{\it Ageing and the glass transition}, Springer (Heidelberg 2007) ({\tt cond-mat/0511743})
\bibitem{Wolf90} D.E. Wolf and J. Villain, Europhys. Lett. {\bf 13}, 389 (1990).
\bibitem{Wright35} E.M. Wright, J. London Math. Soc. {\bf 10}, 287 (1935);
and Proc. London Math. Soc. {\bf 46}, 389 (1940); erratum J. London Math. Soc. {\bf 27}, 256 (1952).
\bibitem{Zinn02} J. Zinn-Justin, {\it Quantum field-theory and critical phenomena}, 4$^{\rm th}$ edition,
Oxford University Press (2002).

\end{thebibliography}
\end{document}